\documentclass[twocolumn,showpacs,aps,prl,superscriptaddress,letterpaper,nofootinbib,10pt]{revtex4-1}

\usepackage{ifpdf}
\ifpdf
\usepackage[colorlinks=true,
 linkcolor=webblue,
 citecolor=webblue,
 bookmarksopen=true,
 bookmarksnumbered=true,
 citebordercolor={.5 .5 .5}]{hyperref}
\hypersetup{%
  pdftitle = {Search for the decay D0->gamma gamma and Measurement of the branching fraction for D0->pi0pi0},
  pdfauthor = {James Morris}
}
\usepackage[pdftex]{graphicx}
\else
\usepackage[dvipdfm]{hyperref}
\usepackage[dvips]{graphicx}
\fi

\usepackage{amsmath}
\usepackage{epsf}
\usepackage{epsfig}
\usepackage{subfigure}
\usepackage{color} 
\usepackage{verbatim} 

\definecolor{webblue}{rgb}{0,0,.8}
\definecolor{ccbarcolor}{rgb}{0.8,0.8,1}
\definecolor{udscolor}{rgb}{0.8,1,0.8}
\definecolor{bpbmcolor}{rgb}{0.8,1,1}
\definecolor{b0b0color}{rgb}{1,1,0.8}

\definecolor{gammagammacolor}{rgb}{1,0.8,1}
\definecolor{pi0pi0color}{rgb}{1, 0.8, 0.6}
\definecolor{kspi0color}{rgb}{1, 0.8, 0.8}
\definecolor{datacolor}{rgb}{0.8, 0.6, 1}

\definecolor{darkgray}{gray}{0.5}

\input ./babarsym

\newcommand{\BABARPubYear}    {11}
\newcommand{\BABARPubNumber}  {04}
\newcommand{\SLACPubNumber} {14392}

\newcommand{\dStar}{D^{*+}}
\newcommand{\dstar}{D^{*+}}
\newcommand{\dStarTag}{D^{*+} \rightarrow D^{0} \pi^{+}}
\newcommand{\dgg}{D^0 \rightarrow \gamma \gamma}
\newcommand{\dpp}{D^0 \rightarrow \pi^0 \pi^0}
\newcommand{\dkp}{D^0 \rightarrow K_S^0 \pi^0}

\newcommand{\bpbm}{B^{+} B^{-}}
\newcommand{\bzbbar}{B^{0} \overline{B}^0}

\newcommand{\dzero}{D^0}

\newcommand{\gammasigeff}{6.1\% \xspace}

\newcommand{\gammayield}{-6 \pm 15}

\newcommand{\gammagammasensitivity}{2.4 \times 10^{-6}}
\newcommand{\gammagammaul}{2.2 \times 10^{-6}}

\newcommand{\pizsigeff}{15.2\% \xspace}

\newcommand{\bfratio}{8.4}
\newcommand{\bfstaterror}{0.1}
\newcommand{\bfsystemerror}{0.4}
\newcommand{\bfreferror}{0.3}

\newcommand{\kspizsigeffforgammacuts}{7.6 \%}
\newcommand{\kspizsigeffforpizcuts}{12.0\%}

\def\dStar{D^{*+}}

\def\dStarTag{\dStar \to D^{0} \piSlow}

\def\dStarTag{D^{*+} \rightarrow D^{0} \pi^{+}}

\def\figurebox#1#2#3{%
    \def\arg{#3}%
    \ifx\arg\empty
    {\hfill\vbox{\hsize#2\hrule\hbox to #2{\vrule\hfill\vbox to #1{\hsize#2\vfill}\vrule}\hrule}\hfill}%
    \else
    {\hfill\epsfbox{#3}\hfill}%
    \fi}
\long\def\inst#1{\par\nobreak\kern 4pt\nobreak
    {\it #1}\par\vskip 10pt plus 3pt minus 3pt}

\begin{document}

\begin{flushleft}
SLAC-PUB-\SLACPubNumber\\
\babar-PUB-\BABARPubYear/\BABARPubNumber\\
\end{flushleft}

\title{ 
{Search for the Decay $D^0 \rightarrow \gamma \gamma$ and Measurement of the Branching Fraction for $D^0 \rightarrow \pi^0 \pi^0$}
}

%
\author{J.~P.~Lees}
\author{V.~Poireau}
\author{E.~Prencipe}
\author{V.~Tisserand}
\affiliation{Laboratoire d'Annecy-le-Vieux de Physique des Particules (LAPP), Universit\'e de Savoie, CNRS/IN2P3,  F-74941 Annecy-Le-Vieux, France}
\author{J.~Garra~Tico}
\author{E.~Grauges}
\affiliation{Universitat de Barcelona, Facultat de Fisica, Departament ECM, E-08028 Barcelona, Spain }
\author{M.~Martinelli$^{ab}$}
\author{D.~A.~Milanes$^{ab}$}
\author{A.~Palano$^{ab}$ }
\author{M.~Pappagallo$^{ab}$ }
\affiliation{INFN Sezione di Bari$^{a}$; Dipartimento di Fisica, Universit\`a di Bari$^{b}$, I-70126 Bari, Italy }
% ...
\collaboration{The \babar\ Collaboration}
\noaffiliation

\begin{abstract}
We search for the rare decay of the $D^{0}$ meson to two photons, $D^0 \rightarrow \gamma \gamma$, and present a measurement of the branching fraction for a $D^0$ meson decaying to two neutral pions, ${B}(D^0 \rightarrow \pi^0 \pi^0)$.  The data sample analyzed corresponds to an integrated luminosity of 470.5 \ensuremath{\mbox{\,fb}^{-1}}\xspace collected by the \mbox{\sl B\hspace{-0.4em} {\small\sl A}\hspace{-0.37em} \sl B\hspace{-0.4em} {\small\sl A\hspace{-0.02em}R}} detector at the PEP-II asymmetric-energy $e^{+} e^{-}$ collider at SLAC.  We place an upper limit on the branching fraction, ${B}(D^0 \rightarrow \gamma \gamma) < 2.2 \times 10^{-6}$, at 90\% confidence level. This limit improves on the existing limit by an order of magnitude. We also find ${B}(D^0 \rightarrow \pi^0 \pi^0) = (8.4 \pm 0.1 \pm 0.4 \pm 0.3)  \times 10^{-4}$.
\end{abstract}



\maketitle

\section{I. INTRODUCTION}\label{sec:intro}

In the Standard Model (SM) flavor-changing neutral currents (FCNC) are forbidden at tree level \cite{PhysRevD.2.1285}. These decays are allowed at higher order and have been measured in the $K$ and $B$ meson systems \cite{RevModPhys.75.1159}. In the charm sector, however, the small mass difference between down-type quarks of the first two families translates to a large suppression at the loop level from the GIM mechanism \cite{PhysRevD.2.1285}. To date, measurements of radiative decays of charm mesons are consistent with results of theoretical calculations that include both short-distance and long-distance contributions and predict decay rates several orders of magnitude below the sensitivity of current experiments \cite{Gaillard:1974hs, Ma:1981eg, Eeg:1990mx, Goity:1986sr, D'Ambrosio:1986ze, Kambor:1993tv}. While these rates are small, it has been postulated that new physics (NP) processes can lead to significant enhancements \cite{Prelovsek:2000xy}. 

In this paper we report results of a search for the FCNC decay of the neutral $D$ meson into two photons.  The only previous study was conducted by the CLEO collaboration using 13.8 \invfb  \cite{Coan:2002te}.  Theoretical calculations predict that the decay $\dgg$ is dominated by long-distance effects.  A calculation in the framework of Vector Meson Dominance (VMD)  \cite{Burdman:2001tf} yields
\begin{equation}
{B}{(D^0 \to\gamma \gamma)}^{\rm (VMD)} \simeq (3.5~^{+4.0}_{-2.6}) \times 10^{-8}.
\end{equation}
\noindent A separate calculation using heavy quark effective theory combined with chiral perturbation theory (HQ$\chi$PT)  \cite{Fajfer:2001ad} reveals a similar dominance of long-distance over short-distance (SD) effects, with the SD branching ratio estimated to be \cite{Burdman:2001tf} 
\begin{equation}
{B}{(D^0\to\gamma\gamma)}^{({\rm SD})} \ \simeq \ 3 \times 10^{-11}.  
\end{equation}

\noindent In the context of the Minimal Supersymmetric Standard Model, gluino exchange can enhance the SM rate by up to a factor of 200 \cite{Prelovsek:2000xy}.  
The large number of charm decays in the $\babar$ dataset provide the opportunity to study this enhancement.  A 200-fold increased rate would result in approximately 1370 events in the $\babar$ dataset (470.5 \invfb) with only six events predicted from the theoretical SM branching fraction ($3.5 \times 10^{-8}$, as determined in the VMD calculation.)  A summary of the relevant branching fractions is shown in  Table \ref{tab:bfulnumbers}.

In this paper we also report a new measurement of the branching fraction for the decay $\dpp$, which is the dominant background in the $\dgg$ analysis. 

\begin{table}[htbp]
  \centering
\caption{Summary of predictions and measured values or limits for branching fractions relevant to this analysis. The results presented in this paper are not included in this table.}
    \begin{tabular*}{\columnwidth}{lcl}
    \hline
\multicolumn{3}{c}{Theoretical predictions}\\
    \hline
    Mode  & Value & Reference \\
    $\dgg$ (SM,VMD) &  $ \approx (3.5~^{+4.0}_{-2.6}) \times 10^{-8}$   & Burdman \cite{Burdman:2001tf}  \\
    $\dgg$ (SM,HQ$\chi$PT) &  $(1.0 \pm 0.5) \times 10^{-8}$   & Fajfer \cite{Fajfer:2001ad}  \\
    $\dgg$ (MSSM) &  $6 \times 10^{-6}$  & Prelovsek \cite{Prelovsek:2000xy}     \\
          &       &  \\
\hline
\multicolumn{3}{c}{Experimental results}       \\
\hline
    Mode  & Value & Reference \\
    $\dgg$  &  $<2.9\times 10^{-5}$  & Coan \cite{Coan:2002te}\\
    $\dpp$ (2006) &  $(7.9 \pm 0.8) \times 10^{-4}$ &  Rubin \cite{Rubin:2005py}\\
    $\dpp$ (2010) &  $(8.1 \pm 0.5) \times 10^{-4}$ &  Mendez \cite{Mendez2010}\\
    $\dkp$ &   $(1.22 \pm 0.05) \times 10^{-2}$    & Nakamura \cite{Nakamura:2010zzi} \\
    \hline
    \end{tabular*}
  \label{tab:bfulnumbers}
\end{table}

\section{II. THE \texorpdfstring{\bf $\babar$}{BaBar} DETECTOR AND DATASET}
 This analysis is based on a data sample corresponding to an integrated luminosity of 470.5 \invfb collected by the $\babar$ detector at the SLAC PEP-II $e^+e^-$ asymmetric-energy collider operating at $\epem$ center-of-mass (CM) energies of $\sqrt{s} = 10.58\gev$ and $10.54 \gev$. 

The \babar\ detector is described in detail elsewhere~\cite{detector}.  Charged particle momenta and positions are measured with a five-layer double-sided silicon  vertex tracker (SVT) and a 40-layer drift chamber (DCH).   Charged hadron identification is provided by measurements of the ionization energy loss, $dE/dx$, in the  tracking system and the Cherenkov angle obtained from  a ring-imaging Cherenkov detector (DIRC).  An  electromagnetic calorimeter (EMC) consisting of 6580  CsI(Tl) crystals measures the energy deposited by electrons and photons.  These detector elements are located inside the cryostat of a superconducting solenoidal magnet, which provides a 1.5 T magnetic field. The instrumented flux return (IFR) of the magnet allows  discrimination of muons from pions.

A detailed Monte Carlo simulation (MC) of the \babar\ detector based on \geant4 \cite{geant4} is used to validate the analysis and determine 
the reconstruction efficiencies.  We use simulated events to optimize the selection criteria by maximizing significance, defined as $N_{S}/\sqrt{N_{S}+N_{B}}$, where  $N_S$ and $N_B$ denote
the number of signal and background candidates in the MC simulation assuming a $\dgg$ branching fraction of $5.4 \times 10^{-6}$ (five times less than the CLEO collaboration upper limit).  The background samples include $\epem \rightarrow \ccbar$, $\epem \rightarrow \qqbar, q = u$, $d$ or $s$, $\epem \rightarrow \bzbbar$, and $\epem \rightarrow \bpbm$ decay modes.

\section{III. EVENT RECONSTRUCTION AND SELECTION}

For the decay modes used in this study we require that the neutral $D$ meson originates in the decay $\dStarTag$, which is referred to as a $\dstar$ tag. (The inclusion of the charge conjugate modes is implied unless otherwise stated.) Without such a tag the signal is dominated by combinatoric background.  To avoid uncertainties in the number of $\dStar$ mesons in the $\babar$ dataset, we perform measurements of the $\dgg$ and $\dpp$ branching fractions relative to a well measured reference mode. The $\dkp$ decay is chosen for this purpose due to its large branching fraction of (1.22 $\pm$ 0.05)\% \cite{Nakamura:2010zzi} and partial cancellation of systematic uncertainties in our measurements.

In the $\dgg$ analysis the $\dzero$ candidate is formed by combining pairs of photon candidates. $\dzero$ candidates are required to have an invariant mass between $1.7$ and  $2.1\gevcc$ for the $\dgg$ analysis and an invariant mass between $1.65$ and $2.05\gevcc$ for the $\dpp$ analysis.  A photon candidate is defined as energy deposited in the EMC, which is not associated with the trajectory of any charged track and which exhibits the appropriate shower characteristics with a lateral moment \cite{Drescher1985464} greater than 0.001. The photon candidates are selected to have CM energies between $0.74$ and $4\gev$.

In the $\dpp$ analysis two $\piz$ candidates each with CM momentum above $0.6\gevc$ are combined to form a $\dzero$ candidate. The $\piz$ candidates are formed by combining two photon candidates with lateral moment less than 0.8. The list of $\piz$ candidates also includes single EMC clusters containing two adjacent photons (merged $\piz$).  

The $\dzero$ candidates for the $\dkp$ reference mode are formed by combining a $\piz$ candidate as defined above with a $K_S^0$ candidate consistent with the decay $K_S^0 \rightarrow \pi^+\pi^-$.  The $\pip\pim$ invariant mass is required to be between $0.491$ and $0.505\gevcc$.  To be selected as a $K_S^0$ candidate the decay length significance must be greater than 3, where the decay length significance is defined as the measured flight length divided by its estimated uncertainty.  

In all modes, $\dzero$ candidates are combined with $\pip$ candidates selected from tracks with CM momentum between $0.05$ and $0.45\gevc$. A kinematic fit is applied to the events, requiring the candidate $\dzero$ invariant mass to be between $1.6$ and $2.1\gevcc$. Both the $\dzero$ and $\pip$ are constrained to originate from a common vertex within the beamspot to satisfy the $\dstar$ tag requirement. 

\section{IV. BACKGROUND STUDIES}\label{sec:background}

Backgrounds from $B$ meson decays are removed by selecting $\dstar$ candidates with CM momentum greater than $2.85 \gevc$ in the case of $\dgg$ and greater than $2.4\gevc$ in the case of $\dpp$. The difference reflects cuts optimized to separate MC samples.

In order to minimize systematic uncertainties the reference mode analysis was performed separately for each of the two signal modes, each time using identical criteria that were optimized for the respective signal mode. These selections result in 95\% rejection of $B$ meson decay modes. The $\dgg$ decay mode has significant backgrounds due to QED processes, which are largely removed by requiring that the total number of charged tracks in the event be greater than four and the number of neutral candidates in the event be greater than four.   

The dominant background to $\dgg$ is due to $\dpp$ decays. To remove this background, we implement a $\pi^0$ veto. From our sample of $\dgg$ candidates we reject all events in which one of the photons can be combined with any other photon candidate in the event to form a $\piz$.  This veto rejects 95\% of the background and keeps 66\% of the signal.

The $\dgg$ analysis signal efficiency is $\gammasigeff$ with the corresponding reference mode ($\dkp, K_S^0 \rightarrow \pip \pim$) efficiency at $\kspizsigeffforgammacuts$. The $\dpp$ analysis signal and reference mode efficiencies are $\pizsigeff$ and $\kspizsigeffforpizcuts$, respectively.

\section{V. FIT PROCEDURE AND RESULTS}\label{sec:fittingprocedure}

For each of the three decay modes we determine the signal yield using unbinned maximum likelihood fits to the invariant mass distribution of $D^{0}$ candidates passing the above selection criteria. The overall probability distribution functions (PDFs) are sums of functions describing signal and background distributions obtained from the Monte Carlo simulation. The relative normalizations of these functions are free parameters while the individual shapes are fixed. 

In the $\dgg$ analysis the signal PDF consists of a Crystal Ball \cite{PhysRevD.34.711} function and a bifurcated Gaussian distribution.  The background PDF is a 2$^{\textrm{nd}}$-order Chebychev polynomial and the $\dpp$ background shape is described by a second Crystal Ball function.  In the $\dpp$ analysis the signal is described by a sum of a Gaussian, a bifurcated Gaussian, and a Crystal Ball function, and a background PDF described by a 3$^{\textrm{rd}}$-order Chebychev polynomial.

\begin{figure}[htbp]
\centering
\includegraphics[width=\columnwidth]{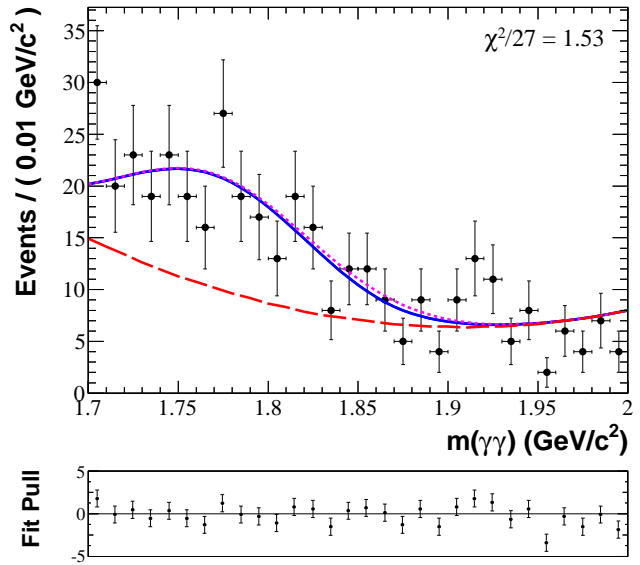}
\caption{The $\gamma\gamma$ mass distribution for $\dgg$ candidates in data (data points). The curves show the result of an unbinned maximum likelihood fit to the measured mass distribution. The solid blue curve corresponds to signal component resulting in a slight negative yield, the long-dash red curve corresponds to combinatoric background component, and the small-dash pink curve corresponds to the combinatoric background plus $\dpp$ background shape. The $\chi^2$ value is determined from binned data and is provided as a goodness-of-fit measure. The pull distribution shows differences between the data and the solid blue curve with values and errors normalized.}
\label{fig:gammafit_data}
\end{figure}

The invariant $\gamma\gamma$ mass distribution obtained from the $\dgg$ analysis is shown in Fig.~\ref{fig:gammafit_data} together with projections of the likelihood fit and the individual signal and background combinations.  The signal yield is $\gammayield$, consistent with no $\dgg$ events. 
We convert this result to a branching fraction for $\dgg$ relative to the $\dkp$ reference mode using
\begin{equation}
{B}(\dgg) = \frac{\frac{1}{\eps_{\gamma\gamma}}N(\dgg)}{\frac{1}{\eps_{K_S^0\pi^0}}N(\dkp)} \times {B}(\dkp),
\label{eq:gammabfcalc}
\end{equation}
\noindent where $N$ and $\eps$ are the yield and efficiency of the respective modes and ${B}(\dkp)$ is the known $\dkp$;$K_S^0 \rightarrow \pip \pim$ branching fraction \cite{Nakamura:2010zzi}.  In this analysis the $\dkp$ signal yield is $126599 \pm 568$ events. We find ${B} (\dgg) = (-0.49 \pm 1.23 \pm 0.02) \times 10^{-6}$ where the errors are the statistical uncertainty and the uncertainty in the reference mode branching fraction, respectively.

\begin{figure}[htbp]
\centering
\includegraphics[width=\columnwidth]{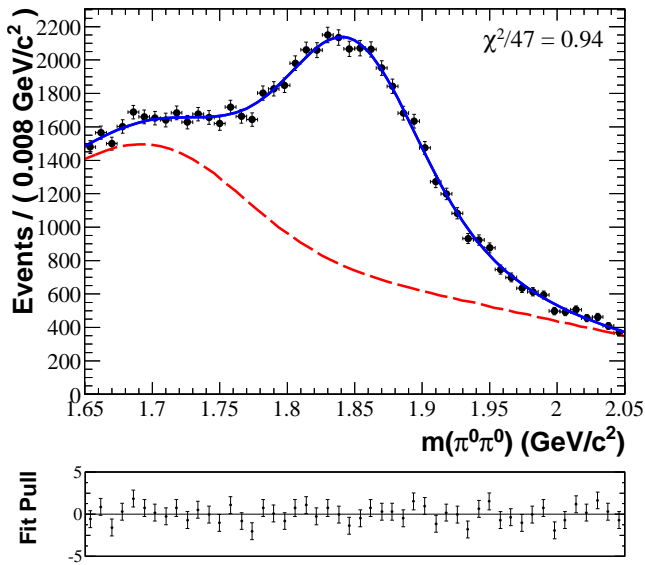}
\caption{The $\piz\piz$ mass distribution for $\dpp$ candidates in data (data points).  The curves show the result of the unbinned maximum likelihood fit to the measured mass distribution. The solid blue curve corresponds to the full PDF including the signal and the dashed red curve corresponds to the combinatoric background component.  The $\chi^2$ value is determined from binned data and is provided as a goodness-of-fit measure. The pull distribution shows differences between the data and the solid blue curve with values and errors normalized.}
\label{fig:pi0fit_data}
\end{figure}

The invariant mass distribution for events in the $\dpp$ analysis is shown in Fig.~\ref{fig:pi0fit_data}.  The signal yield 
for $\dpp$ is 26010 $\pm$ 304 events. For $\dkp$ (mass distribution not shown) the signal yield is 207538 $\pm$ 1143 events. Adjusting
Eq. \ref{eq:gammabfcalc} for the $\dpp$ case we convert this yield to a branching fraction and find 
${B} (\dpp) = (8.4 \pm 0.1 \pm 0.3) \times 10^{-4}$. The first error denotes the statistical uncertainty and the second error reflects the uncertainties in the reference mode branching fraction.

\section{VI. SYSTEMATIC UNCERTAINTIES}\label{sec:systematics}

Several systematic uncertainties cancel partially or completely when the branching fraction is measured with respect to the $\dkp$ 
reference mode. The uncertainty in tracking efficiency and vertexing 1.39\%. The uncertainty due to photon reconstruction efficiency in 
the ratio of the signal mode branching fraction to the reference mode branching fraction is 3.0\% and 0.6\% for the $\dpp$ and $\dgg$ 
analyses, respectively.

In order to account for the uncertainty arising from fixed PDF shapes, the parameters determined from the Monte Carlo simulation, are varied by random amounts sampled from the covariance matrix retaining correlations among parameters.  The values of these parameters are fixed and the resulting PDF is fit to data allowing the yield to float, the 1$\sigma$ width of the obtained signal yield distribution is taken as the systematic uncertainty.   In the $\dpp$ analysis, fixing the signal and combinatoric background shapes results in 0.20\% and 0.80\% systematic uncertainties, respectively. Fixing the $\dkp$ signal and background shapes for the reference mode results in 0.17\% and 0.63\% systematic uncertainties, respectively. 

Potential differences in $\piz$ veto efficiencies between data and the Monte Carlo simulation are estimated using a sample of candidates for the physically forbidden decay $D^{0} \rightarrow K_{S}^{0} \gamma$. The difference in the ratios of numbers of candidates before and after the veto between data and the Monte Carlo simulation is taken as the systematic uncertainty.  We measure the difference as a function of the number of photons in the event and as a function of the photon energy. In all cases,
 the variations are found to be less than or equal to 1.8\%.

In order to account for imperfect modeling of $\dstar$ hadronization, a 4\% correction is applied to the MC for normalized momenta, $x = p(\dstar)/p_{\rm max}(\dstar)$, within the region $x = 0.575$ to $x = 0.7$ to match cross-section measurements made by the CLEO collaboration \cite{Colangelo1992167}.  We calculate the ratios of signal efficiencies ($\eps_{\dgg}/\eps_{\dkp}$, $\eps_{\dpp}/\eps_{\dkp}$) with and without this correction applied to the MC and determine systematic uncertainties of 0.02\% and 0.03\% for the $\dgg$ and $\dpp$ modes, respectively, due to this correction.

To account for systematic uncertainties due to applying a particular set of selection criteria, we vary the selection criteria and recalculate the results. The reconstruction efficiency is determined from MC and the efficiency-corrected yield is measured from data when each set of selection criteria is applied.  These yields are found to be distributed normally and the standard deviation is taken to be the systematic uncertainty.  Choosing particular event selections for the $\dpp$ and $\dkp$ studies results in systematic uncertainties of 2.50\% and 0.76\%, respectively.  

The systematic uncertainties are summarized in Table \ref{tab:sysSummary}.  For the $\dpp$ mode a total systematic uncertainty of $4.2\%$ 
is obtained by adding all contributions in quadrature.

\begin{table}[htbp]                                                            
  \begin{center}       
\caption{Summary of  systematic uncertainties, $\sigma$, for each mode. [*] These sources of systematic uncertainty are assessed in a
combined Monte Carlo simulation study. See text for details.}\label{tab:sysSummary}                                                     
    \begin{tabular}{lcc} 
	\hline                                                                                                              
      Source of  & $ \sigma(\dgg)$ & $\sigma(\dpp)$ \\
      Systematic Uncertanity & (\%) & (\%) \\
  \hline 
Tracking ($K_S^0$) and Vertexing & 1.39 & 1.39 \\ 
Photon Reconstruction & 0.60 & 3.0 \\
$\pi^{0}$ Veto & 1.8 & - \\
$\dstar$ Hadronization & 0.02 & 0.03 \\
Signal  Shape & * & 0.20 \\
Background Shape & * & 0.80  \\
Selection Criteria & * & 2.5 \\
$\dkp$ Signal Shape & 0.10 & 0.17 \\
$\dkp$ Background Shape & 0.53 & 0.63 \\
$\dkp$ Selection Criteria & 0.76 & 0.76 \\
\hline
      Total Systematic Uncertainty & $ * $ & $4.3$ \\
      \hline
    \end{tabular}
  \end{center}
\end{table}

\begin{figure}[htbp]
\centering
\includegraphics[width=\columnwidth]{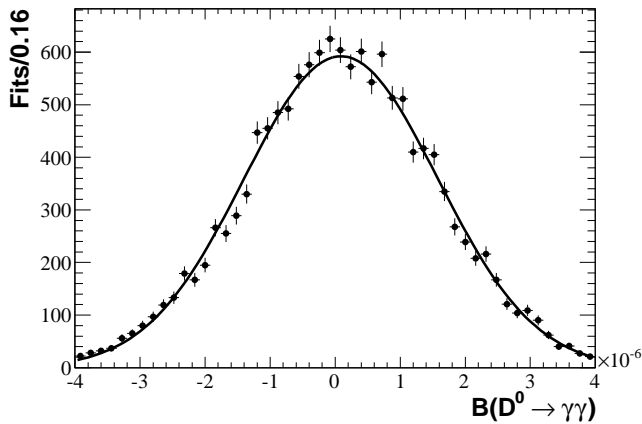}
\caption{Distribution of branching fraction calculations when varying all sources of uncertainty. }
\label{fig:upperlimitdistro}
\end{figure}

For the $\dgg$ analysis we combine all systematic uncertainties with the statistical uncertainties in the upper-limit 
calculation. 
In a Monte Carlo simulation study we generate event samples using the complete background PDF from the data fit and repeat the 
branching fraction calculation 14000 times varying all sources of systematic uncertainties in the process.
For each branching fraction calculation the selection values on the continuous variables are varied within ranges established 
from the $\dpp$ analysis. In each calculation the parameters of the signal and background PDFs are varied within their uncertainties
while fully accounting for the correlations among them. Systematic uncertainties such as tracking, photon reconstruction, $\piz$ veto, 
and $\dstar$ hadronization are added in quadrature and the $\dgg$ signal efficiency is varied randomly according to a normal 
distribution with a width equal to this total uncertainty.  To account for the uncertainty in the branching fraction of the $\dkp$ reference mode, the nominal value is varied randomly according to a 
normal distribution with a width equal to the established $\dkp$ branching fraction uncertainty \cite{Nakamura:2010zzi}.
The resulting distribution of ${B} (\dgg)$ branching fractions is shown in 
Fig.~\ref{fig:upperlimitdistro}.

Integrating this distribution to $90\%$ of the area with
${B}(\dgg)>0$ gives us the expected sensitivity of our analysis. We find
\begin{equation}
{B}(\dgg) < \gammagammasensitivity
\end{equation}
at $90\%$ confidence level.

\section{VII. RESULTS}\label{sec:results}

In this paper we present a new measurement for the branching fraction for a $D^0$ meson decaying to two neutral pions:

\begin{equation}
{B}(\dpp) = (\bfratio \pm \bfstaterror \pm \bfsystemerror \pm \bfreferror) \times 10^{-4},
\end{equation}

\noindent where the errors denote the statistical, systematic, and reference mode branching fraction uncertainties, respectively.

We also report the result of a search for the decay of a neutral $D$ meson to two photons. The observed yield is $-6 \pm 15$ consistent
with no $\dgg$ events. Our analysis has an expected sensitivity of ${B} (\dgg) < \gammagammasensitivity$ at $90\%$ confidence level.
In order to obtain an upper limit for ${B} (\dgg)$ we repeat the sensitivity study described in Section~VI with the
signal yield set to the measured value of $-6$ instead of $0$ and find
\begin{equation}
{B}(\dgg) < \gammagammaul
\end{equation}

\noindent at 90\% confidence level.

This result is consistent with our expected sensitivity and with SM expectations. 
As stated earlier, gluino exchange has been postulated to possibly enhance the SM rate by up to a 
factor of 200 \cite{Prelovsek:2000xy}. Based on the upper limit of the branching fraction for $\dgg$ presented 
in this paper, the enhancement of the rate over the expected SM rate cannot exceed a factor of 70.

\section{IX ACKNOWLEDGMENTS}\label{sec:acknowledgements}

We are grateful for the 
extraordinary contributions of our \pep2\ colleagues in
achieving the excellent luminosity and machine conditions
that have made this work possible.
The success of this project also relies critically on the 
expertise and dedication of the computing organizations that 
support \babar.
The collaborating institutions wish to thank 
SLAC for its support and the kind hospitality extended to them. 
This work is supported by the
US Department of Energy
and National Science Foundation, the
Natural Sciences and Engineering Research Council (Canada),
the Commissariat \`a l'Energie Atomique and
Institut National de Physique Nucl\'eaire et de Physique des Particules
(France), the
Bundesministerium f\"ur Bildung und Forschung and
Deutsche Forschungsgemeinschaft
(Germany), the
Istituto Nazionale di Fisica Nucleare (Italy),
the Foundation for Fundamental Research on Matter (The Netherlands),
the Research Council of Norway, the
Ministry of Education and Science of the Russian Federation, 
Ministerio de Ciencia e Innovaci\'on (Spain), and the
Science and Technology Facilities Council (United Kingdom).
Individuals have received support from 
the Marie-Curie IEF program (European Union), the A. P. Sloan Foundation (USA) 
and the Binational Science Foundation (USA-Israel).


%
\end{document}